\documentstyle[12pt]{article}
\begin{document}
\title{A Note on "Extension, Spin and Non Commutativity"}
\author{B.G.Sidharth \\
Centr for Applicable Mathematics and Computer Sciences\\
B.M.Birla Science Centre,Hyderabad, 500463,India}
\date{}
\maketitle
\begin{abstract}
We show that the Dirac theory of the electron, corresponds to recent approaches
based on a Non commutative spacetime.
\end{abstract}
\section{Introduction}
As is well known, in Quantum Theory, as we go down to arbitrarily small spacetime
intervals, we begin to encounter arbitrarily large momenta and energy. This
in Quantum Field Theory manifests itself in the form of divergences. Indeed
in the Dirac Theory of the electron \cite{r1} the position operator takes on
complex eigen values and it is only after an averaging over intervals at the
Compton scale that we recover usual physics. This non Hermiticity of the position
operator has been recognised for a long time \cite{r2,r3}, and
meaningful position operators have been constructed. The crux of the matter,
in this latter approach, as also in the Foldy-Wolthysen transformation, which
achieves the same purpose \cite{r3,r4}, is that operators are constructed
in such a way that the so called positive energy spinor and negative energy
spinor of the Dirac bi-spinor wave function, do not mix. This is quite meaningful
because the negative energy spinors are negligible outside the Compton wavelength
region. Indeed, Dirac's averaging over the Compton scale, referred to, achieves
the same purpose-- it eliminates zitterbewegung effects.\\
The spirit of Dirac's average spacetime intervals rather than spacetime points
has continued to receive attention over the years in the form of minimum
spacetime intervals-- from the work of Snyder and Schild to Quantum Superstring
theory \cite{r5,r6,r7,r8,r9,r10,r11}. In modern
language, it is symptomatic of a Non commutative spacetime geometry \cite{r12,r13,r14}.
We will now highlight the intimate connection between spin, non commutativity and extension
in the above context.
\section{The Non commutative Structure}
Indeed Newton and Wigner showed that the correct physical coordinate operator
is given by
\begin{equation}
x^k = (1 +\gamma^0 ) \frac{p_0^{3/2}}{(p_0+\mu)^{1/2}} \left(-\frac{\imath \partial}
{\partial p_k}\right) \frac{p_0^{-1/2}}{(p_0+\mu)^{1/2}} E\label{e1}
\end{equation}
where $E$ is a projection operator given by
$$
E = \frac{1}{2} p_0(E\gamma^k p_k + \mu )\gamma^0$$
and the gammas denote the usual Dirac matrices.\\
To appreciate the significance of (\ref{e1}), we first consider the case of
spin zero. Then (\ref{e1}) goes over to
\begin{equation}
x^k = \imath \frac{\partial}{\partial p_k} + \frac{1}{8\pi} \int
\frac{\exp (-\mu |(x-y|)}{|x-y|} \frac{\partial}{\partial y} dy\label{e3}
\end{equation}
The first term on the right side of (\ref{e3}) denotes the usual position operator,
but the second term represents an imaginary part, which has an extension $\sim 1/\mu$,
the Compton wavelength, exactly as in the case of the Dirac electron.\\
Returning to Dirac's treatment \cite{r1}, the position coordinate is given by
\begin{equation}
\vec x = \frac{c^2pt}{H} + \frac{1}{2} \imath c \hbar (\vec \alpha - c\vec p
H^{-1}) H^{-1} \equiv \frac{c^2p}{H}t + \hat x\label{e4}
\end{equation}
$H$ being the Hamiltonian operator and $\alpha$'s the non-commuting Dirac
matrices, given by
$$\vec \alpha = \left[\begin{array}{l}
\vec \sigma \quad 0\\
0 \quad \vec \sigma
\end{array}
\right]
$$
The first term on the right hand side is the usual (Hermitian) position. The
second term of $\vec x$ is the small oscillatory term of the order of the Compton
wavelength, arising out of zitterbewegung effects which averages out to zero.
On the other hand, if we were to work with the (non Hermitian) position operator in
(\ref{e4}), then we can easily verify that the following Non-commutative geometry
holds,
\begin{equation}
[x_\imath , x_j] = \alpha_{\imath j} l^2\label{e5}
\end{equation}
where $\alpha_{\imath j} \sim 0(1)$.\\
The relation (\ref{e5}) shows on comparison with the position-momentum commutator
that the coordinate $\vec x$ also behaves like a "momentum". This can be
seen directly from the Dirac theory itself where we have
\begin{equation}
c\vec \alpha = \frac{c^2 \vec p}{H} - \frac{2\imath}{\hbar} \hat x H\label{e6}
\end{equation}
In (\ref{e6}), the first term is the usual momentum. The second term is the extra
"momentum" $\vec {\hat p}$ due to the relations (\ref{e5}).\\
Infact we can easily verify from (\ref{e6}) that
\begin{equation}
\vec {\hat p} = \frac{H^2}{\hbar c^2} \hat x\label{e7}
\end{equation}
where $\hat x$ has been defined in (\ref{e4}).\\
We finally investigate what the angular momentum $\sim \vec x \times \vec p$ gives -
that is, the angular momentum at the Compton scale. Using (\ref{e4}), we can
easily show that
\begin{equation}
(\vec x \times \vec p )_z = \frac{c}{E} (\vec \alpha \times \vec p )_z = \frac{c}{E}
(p_2 \alpha_1 - p_1\alpha_2)\label{e8}
\end{equation}
where $E$ is the eigen value of the Hamiltonian operator $H$. (\ref{e8})
shows that the angular momentum leads to the "mysterious" Quantum Mechanical spin.\\
In the above considerations, we started with the Dirac equation and deduced
the underlying Noncommutative geometry of spacetime. Interestingly, starting
with Snyder's Non commutative geometry, based solely on Lorentz invariance and a
minimum spacetime length, at the Compton scale,
$$[x,y] = \frac{\imath l^2}{\hbar} L_z etc.$$
that is, in effect starting with (\ref{e5}), it is possible to deduce the relations
(\ref{e8}),(\ref{e7}) and the Dirac equation itself \cite{r10,r15,r16,r17}.\\
We have thus established the correspondence between considerations starting from
the Dirac theory of the electron and Snyder's (and subsequent) approaches based
on a minimum spacetime interval and Lorentz covariance.
\section{Concluding Remarks}
We remark that in the usual Quantum Field Theory we encounter divergences
which require renormalization. The motivation for the extension of particles,
or the Non commutative structure (\ref{e5}) has been to circumvent these divergences.
In this sense the Non commutative geometry represents "renormalised" coordinates.\\
We consider in a little more detail\cite{r17} the implications of Dirac's
averaging over the Compton scale.\\
We consider for simplicity, the free particle Dirac equation. The
solutions are of the type,
\begin{equation}
\psi = \psi_A + \psi_S\label{e25}
\end{equation}
where
$$
\psi_A =   e^{\frac{\imath}{\hbar} Et} \ \left(\begin{array}{l}
                                          0 \\ 0 \\ 1 \\ 0
                             \end{array}\right) \mbox{ or } \ e^{\frac{\imath}{\hbar} Et}
                          \ \   \left(\begin{array}{l}
                                 0 \\ 0 \\ 0 \\ 1
                              \end{array}\right) \mbox{ and }
$$
\begin{equation}
\label{e26}
\end{equation}
$$
\psi_S =  e^{-\frac{\imath}{\hbar} Et} \ \left(\begin{array}{l}
                                 1 \\ 0 \\ 0 \\ 0
                               \end{array}\right) \mbox{ or } e^{-\frac{\imath}{\hbar} Et}
                            \ \   \left(\begin{array}{l}
                                 0 \\ 1 \\ 0 \\ 0
                               \end{array}\right)
$$
denote respectively the negative energy and positive energy solutions. From
(\ref{e25}) the probability of finding the particle in a small volume
about a given point is given by
\begin{equation}
| \psi_A + \psi_S|^2 = |\psi_A|^2 + |\psi_S|^2 +
(\psi_A \psi_S^* + \psi_S \psi_A^*)\label{e27}
\end{equation}
Equations (\ref{e26}) and (\ref{e27}) show that the negative energy and
positive energy solutions form a coherent Hilbert space and so the
possibility of transition to negative energy states exists. This difficulty
however can be overcome by the well known Hole theory which uses the Pauli exclusion
principle, and is described in many standard books on Quantum Mechanics.\\
However the last or interference term on the right side of (\ref{e27}) is like the
zitterbewegung term. When we remember that we really have to consider
averages over space time intervals of the order of $\hbar/mc$ and
$\hbar/mc^2$, this term disappears and effectively the negative energy
solutions and positive energy solutions stand decoupled in what is now
the physical universe.\\
A more precise way of looking at this is\cite{r3} that as is well known, for
the homogeneous Lorentz group, $\frac{p_0}{|p_0|}$ commutes with all operators
and yet it is not a multiple of the identity as one would expect according
to Schur's lemma: The operator has the eigen values $\pm 1$ corresponding
to positive and negative energy solutions. This is a super selection
principle \cite{r18} pointing to the two incoherent Hilbert spaces or universes
now represented by states $\psi_A$ and $\psi_S$ which have
been decoupled owing to the averaging over the Compton scale spacetime
intervals. But the energies we usually consider are such that our length scale
is much greater than the Compton wavelength-- as if we were in the usual poiint
spacetime manifold.\\
Thus once again we see that outside the Compton scale region we
recover the usual physics.\\
It may be mentioned that in Quantum Superstring theory also we encounter
Non commutative relations like (\ref{e4})\cite{r19}.\\
It may also be mentioned that Zakruzewski has shown from a classical viewpoint
that spin implies Non commutative spacetime\cite{r20}.\\
We finally make the following comment:\\
It can easily be verified that if we specialise to the one dimensional case, then
the Dirac coordinate (\ref{e4}) becomes identical to the Newton-Wigner
coordinate, which latter however defines a commutative spacetime in three
dimensions (Cf.ref.\cite{r3}). Further the Nelson-Nottale approach which uses
a double Weiner stochastic process leads to a complex coordinate on the one hand,
and the Schrodinger equation on the other (Cf.ref.\cite{r17} and \cite{r21}).
Such a complex coordinate also appears in the de Broglie-Bohm approach (Cf.ref.\cite{r17}
for details). On the other hand in the Dirac coordinate (\ref{e4}), this
additional coordinate is directly related to zitterbewegung effects. We can see
that the double Weiner process alluded is also connected with such an effect:
The negative time derivative, which does not equal the positive time derivative
in this case, represents negative energy states.\\
However if we analyse the stochastic Schrodinger equation question further,
and generalise the one dimensional case we consider to three dimensions, then
as is well known\cite{r22},
$$\imath \to \vec \sigma ,$$
where $\vec \sigma$ are the Pauli matrices. In other words, we not only cross
over to special relativity, but also recover the non commutative geometry
(\ref{e5}) or Snyder's relations alluded to, at the Compton scale. Thence
it is possible to derive the Dirac equation itself (Cf.ref.\cite{r17}).\\
In other words non commutative spacetime and spin can be shown to originate
from stochastic double Weiner processes at the micro level.


\begin{thebibliography}{99}
\bibitem {r1} P.A.M. Dirac, "The Principles of Quantum Mechanics", Clarendon
Press, Oxford, 1958, pp.4ff, pp.253ff.
\bibitem {r2} T.D. Newton, and E.P. Wigner, Rev. Mod. Phys., \underline{21}
(3), 1949, pp.400ff.
\bibitem {r3} S.S. Schweber, "Relativistic Quantum Field Theory", Harper
and Row, New York, 1964, p.47.
\bibitem {r4} J.D. Bjorken and S.D. Drell, "Relativisitic Quantum Mechanics",
Mc-Graw Hill, New York, 1964, p.24.
\bibitem {r5} A. Schild, Phys.Rev., 73, 1948, p.414-415.
\bibitem {r6} H.S. Snyder, Physical Review, Vol.72, No.1, July 1 1947, p.68-71.
\bibitem {r7} V.G. Kadyshevskii, J.Exptl.Theoret.Phys., USSR, 41, December
1961, p.1885-1894.
\bibitem {r8} C. Wolf, Hadronic Journal, Vol.13, 1990, p.22-29.
\bibitem {r9} T.D. Lee, "Particle Physics and Introduction to Field Theory",
Harwood Academic, 1981, pp.383ff.
\bibitem {r10} B.G. Sidharth, Chaos, Solitons and Fractals, 11(8), 2000, 1269-1278.
\bibitem {r11} B.G. Sidharth, Chaos, Solitons and Fractals, 13(2), 2002, p.189-193.
\bibitem {r12} J. Madore, Class.Quantum Grav. 9, 1992, p.69-87.
\bibitem {r13} B.G. Sidharth, "A Reconciliation of Electromagnetism and
Gravitation", to appear in Annales de la Fondation Louis d Broglie.
\bibitem {r14} B.G. Sidharth, Il Nuovo Cimento, 116B (6), 2001, pg.4 ff.
\bibitem {r15} B.G. Sidharth, Frontiers of Fundamental Physics 4, Plenum
Publishers/Kluwer Academic, New York, 2001, p.97-107.
\bibitem {r16} B.G. Sidharth, "Spin and Non Commutative Geometry", to appear
in Chaos, Solitons and Fractals.
\bibitem {r17} B.G. Sidharth, "Chaotic Universe: From the Planck to the Hubble Scale",
Nova Science Publishers, New York, 2001, p.20.
\bibitem {r18} P. Roman, "Advanced Quantum Theory", Addison-Wesley,
Reading, Mass, 1965, p.31.
\bibitem {r19} Y. Ne'eman, in Proceedings of the First Internatioinal Symposium,
"Frontiers of Fundmental Physics", Eds. B.G. Sidharth and A. Burinskii,
Universities Press, Hyderabad, 1999, p.83-96.
\bibitem {r20} S. Zakruzewski in "Quantization, Coherent States and Complex
Structures", Eds. J.P. Antoine et al., Plenum Press, New York, 1995,
pp.249-255.
\bibitem {r21} L. Nottale, "Fractal Space-Time and Microphysics: Towards
a Theory of Scale Relativity", World Scientific, Singapore, 1993, p.145ff.
\bibitem {r22} M. Sachs, "General Relativity and Matter", D. Reidel
Publishing Company, Holland, 1982, p.45ff.


\end{thebibliography}
\end{document}